\documentclass[twocolumn,showpacs,preprintnumbers,amsmath,amssymb]{revtex4}
\small
\usepackage{amsfonts,amssymb,latexsym,amsmath,amscd}
\usepackage{times,mathptm}

\newcommand{\ket}[1]{|#1\rangle}

\newcommand{\tr}{\mathrm{tr}}

\newcommand{\ff}{\gamma}

\newenvironment{example}
    {
    \smallskip
    \refstepcounter{theorem}
    \noindent
    {\bf Example \Roman{section}.\arabic{theorem}} \ \ }
    {\hspace*{\fill}{$\Diamond$}
                  \smallskip}

    {
    \smallskip
    \refstepcounter{theorem}
    \noindent
    {\bf Remark \Roman{section}.\arabic{theorem}} \ \ }
    {\hspace*{\fill}{$\Diamond$}
    \smallskip}

    {
    \smallskip
    \refstepcounter{theorem}
    \noindent
    {\bf Definition \Roman{section}.\arabic{theorem}} \ \ }
    {\hspace*{\fill}{\ }
    \smallskip}

\newenvironment{proof}[1][]
    {
    \noindent
    {\bf Proof{#1}:  }
    }
    {\hspace*{\fill}{$\Box$}\smallskip}

    {
    \noindent
    {\bf Sketch{#1}:  }
    }
    {\hspace*{\fill}{$\Box$}\smallskip}

    {
    \noindent
    {\bf Reference:  }
    }
    {\hspace*{\fill}{$\odot$}\smallskip}

\newtheorem{theorem}{Theorem}[section]
\newtheorem{proposition}[theorem]{Proposition}

\newcommand{\botCNOT}{
\begin{picture}(0,0)
    \put(1,3){\circle{1}}
    \put(0,1){\line(1,0){2.3}}
    \put(1,1){\line(0,1){2.5}}
    \put(1,1){\circle*{0.4}}
    \put(0,3){\line(1,0){2.3}}
\end{picture}
}

\newcommand{\topCNOT}{
\begin{picture}(0,0)
    \put(1,3){\circle*{0.4}}
    \put(0,3){\line(1,0){2.3}}

    \put(1,1){\circle{1}}
    \put(1,0.5){\line(0,1){2.5}}
    \put(0,1){\line(1,0){2.3}}
\end{picture}
}

\newcommand{\boxGate}[1]{
\begin{picture}(0,0)
    \put(0.5,0.5){\framebox(1,1){\small #1}}
    \put(0,1){\line(1,0){0.5}}
    \put(1.5,1){\line(1,0){0.5}}
\end{picture}
}

\newcommand{\hWire}{
   \begin{picture}(0,0)
    \put(0,1){\line(1,0){2}}
   \end{picture}
}

\begin{document}


\title{Recognizing Small-Circuit Structure in Two-Qubit Operators\\
       and Timing Hamiltonians to Compute Controlled-Not Gates}

\email{vshende@umich.edu}
\author{Vivek V. Shende$^1$,Stephen S. Bullock$^2$
       and Igor L. Markov$^3$}

\email{stephen.bullock@nist.gov}

\affiliation{$^1$The University of Michigan,Department of Mathematics,\\
$^2$National Institute of Standards and Technology, I.T.L.-M.C.S.D.\\
$^3$The University of Michigan,
Department of Electrical Engineering and Computer Science}
\email{imarkov@umich.edu}


\begin{abstract}

  This work proposes numerical tests which determine whether a two-qubit
  operator has an atypically simple quantum circuit.
  Specifically, we describe formulae, written in terms of matrix coefficients,
  characterizing operators implementable with exactly zero, one, or two
  controlled-not ({\tt CNOT}) gates and all other gates being one-qubit.
  We give an algorithm for synthesizing two-qubit circuits with
  optimal number of {\tt CNOT} gates, and illustrate it on
  operators appearing in quantum algorithms by Deutsch-Josza, Shor
  and Grover. In another application, our explicit numerical tests allow timing
  a given Hamiltonian to compute a {\tt CNOT} modulo one-qubit gates, 
  when this is possible.

\end{abstract}

\maketitle


\section{Introduction}

  Quantum circuits compactly represent unitary operators and find
  applications in quantum computing, communication and cryptography
  \cite{NielsenC:00}.  Such a representation can often be
  interpreted as a program (e.g., a sequence of RF pulses for NMR)
  whose execution on a quantum system of choice performs a requested
  unitary evolution. Simple steps in the program correspond to gates
  in the circuit, and smaller circuits lead to faster programs.  In
  this work we discuss exact implementations of two-qubit operators
  because (i) such operators suffice to implement arbitrary
  operators \cite{DiVincenzo:95}, and (ii) a number of controllable
  two-qubit systems were recently reported.

  The simulation of generic two-qubit operators via {\tt CNOT}
  gates and one-qubit operators has been thoroughly investigated,
  resulting in several three-{\tt CNOT} decompositions
  \cite{VidalDawson:03, VatanWilliams:03, ShendeEtAl:04}. It is known 
  that the swap gate requires three {\tt CNOT}s \cite{VatanWilliams:03}, 
  and also that an arbitrary $n$-qubit operator requires at least
  $\lceil \frac{1}{4}(4^n - 3n -1)\rceil$. The proof of this
  latter result \cite{ShendeEtAl:04} holds for any controlled-$u$ gate,
  where $u$ is a given fixed one-qubit operator. For $n=2$, it has been
  shown that an arbitrary controlled-$u$ gate is generically worse
  than the {\tt CNOT} \cite{Zhang+3:03}.

  The above-mentioned results motivate the focus on the {\em basic-gate} 
  library \cite{BarencoEtAl:95}, which consists
  of the {\tt CNOT} gate and all one-qubit gates: it is powerful
  and well-understood. Yet, given the diversity of implementation
  technologies, it is not clear that the {\tt CNOT} gate will
  be directly available in a given implementation. Nonetheless, we believe
  results expressed in the {\em basic-gate} library will be
  relevant. An analogous situation occurs in the design of (classical) 
  integrated
  circuits. In this context, first {\em technology-independent synthesis} is
  performed in terms of abstract gates (AND, OR, NOT). Later, during {\em
  technology mapping}, circuits are converted to use gates that are specific to
  a given implementation technology (e.g., NOR, NAND and AOI gates, which
  require very few {\tt CMOS} transistors). Work in the direction
  of quantum technology mapping includes techniques for expressing
  a {\tt CNOT} gate in terms of a given entangling two-qubit
  gate and arbitrary one-qubit gates \cite{BremnerEtAl:02a}. 
  The simulation of {\tt CNOT} gates with implementation-specific
  resources is the basis of a major physical implementation
  technology \cite{WinelandEtAl:98}.

  The analogy with classical logic synthesis provides the
  following additional intuition: operators useful in practice
  will not be the worst-case operators studied in the
  aforementioned works. This belief is confirmed by published quantum
  algorithms and communication protocols. It is therefore
  important for quantum logic synthesis techniques to detect when
  a given operator can be implemented using fewer gates than are
  necessary in the worst case. For some classes of operators, this
  is easy; e.g., the algorithm in \cite{BullockM:03}
  implements tensor-product operators without {\tt CNOT}s.
  The matrix of a controlled-$U$ operator
  can be recognized by its pattern of zeros and ones
  (either directly, or after pre- and post-multiplication
   by wire swaps).
  Song and Klappenecker \cite{SongK:03} study optimal implementations
  of two-qubit controlled-unitary operators, known to require up to two
  {\tt CNOT} gates. They contribute a catalog of numerical tests
  that detect when zero, one or two {\tt CNOT} gates are required,
  and similar criteria for the number of basic one-qubit gates.

  We address a related question for arbitrary two-qubit operators
  and contribute simple numerical tests to determine
  the minimal achievable number of {\tt CNOT}s, including a novel
  one-{\tt CNOT} test. We also generalize a two-{\tt CNOT} test
  from \cite{VidalDawson:03} and make it easier to compute.
  Such explicit numerical tests facilitate a new application.
  A given two-qubit Hamiltonian $H$, if timed precisely, may allow
  one to implement a {\tt CNOT} using $\mbox{e}^{iHt}$ and one-qubit gates.
  We show how to compute correct durations.

\section{Background and Notation}
\label{sec:background}

It is well known that an arbitrary one-qubit gate $u$ can be
written as $u =\mbox{e}^{i\Phi} R_z(\theta) R_y(\phi) R_z(\psi)$
\cite{NielsenC:00}. Furthermore, the Bloch sphere isomorphism
suggests that the choice of $y, z$ is arbitrary in the sense that
any pair of orthogonal vectors will do: in particular, we may
write \[u =\mbox{e}^{i\Phi} R_z(\theta) R_x(\phi) R_z(\psi) =
\mbox{e}^{i\Phi} R_x(\alpha) R_z(\delta) R_x(\beta)\] These
decompositions are more convenient when working with {\tt CNOT}
gates because $R_z$ gates commute through the control of the {\tt
CNOT} whereas $R_x$ gates commute through the target. We will
denote by $C_j^k$ a {\tt CNOT} with control on the $j$-th wire and
target on the $k$-th. For convenience, we consider the {\tt CNOT}
gate to be normalized to have determinant $1$.

Additional conventions are as follows.  For $g$
any complex matrix, $g^t$ denotes the transpose and $g^*$
denotes the adjoint, i.e. the complex-conjugate transpose.  Additionally,
$\chi(g)=p(x)=\mbox{det }( x I - g)$ denotes the characteristic
polynomial of $g$.  We use axis-dependent phase operators \cite{ShendeEtAl:04}
$S_* = R_*(\pi/2)$, $*=x,y,z$.
Finally, $SU(4)$ denotes the group of all
\emph{determinant one} unitary matrices, fixing the global phase
of a two-qubit unitary operator up to $\pm 1, \pm i$.

We now consider when two-qubit operators $u,v$ differ by pre- or
post-composing with one-qubit operators and possibly by an
irrelevant global phase. In this case, we write $u \equiv v$ and
say that $u$ and $v$ are equivalent up to one-qubit gates. The
following invariant characterizes when this occurs.

\begin{proposition} \label{prop:invariants}
Let $\ff: U(4) \to U(4)$ be given by the formula $u \mapsto u
(\sigma^y)^{\otimes 2} u^t (\sigma^y)^{\otimes 2}$. Then for $u, v
\in SU(4)$, $u \equiv v \iff \chi[\gamma(u)] = \chi[\pm
\gamma(v)]$.
\end{proposition}

We defer the proof to the Appendix.  However, note that this proof
provides an explicit procedure for computing the one-qubit
operators $a,b,c,d \in SU(2)$ such that $(a \otimes b) u (c
\otimes d) = e^{i \phi} v$ in the event that $\chi[\gamma(u)] =
\chi[\pm\gamma(v)]$. We discuss $\ff$ more fully in the context of
minimal universal two-qubit circuits \cite{ShendeEtAl:04}. Related
invariants are discussed in \cite{ZhangEtAl:02,Makhlin:00}, and
generalizations in \cite{BullockBrennen:03}.

\section{Optimizing {\tt CNOT}-count}\label{sec:CNOT}

We now characterize which two-qubit operators admit
a quantum circuit using only $m$ {\tt CNOT} gates. Since
any two-qubit operator is implemented by some three {\tt CNOT}
circuit, the relevant cases are $m=0,1,2$. We begin with case
$m=0$.

\begin{proposition}
  \label{prop:cnotcount:0}
  An operator $u \in SU(4)$ can be simulated using no {\tt CNOT} gates
  and arbitrary one-qubit gates from $SU(2)$ iff $\chi[\ff(u)] =
  (x+1)^4$ or $(x-1)^4$.
\end{proposition}

\begin{proof}
  $u$ can be simulated using no {\tt CNOT} gates iff $u \equiv I$. Thus
  $\chi[\ff(u)] = \chi[\pm\ff(I)] = \chi[\pm I] = (x \pm 1)^4$.
\end{proof}

The case $m=1$ is similar.  
Note that this test \emph{requires} normalizing the global phase
so that $\mbox{det}(v)=1$, implicit in $v \in SU(4)$.  Had we not
normalized the {\tt CNOT} gate, $\chi[\gamma(C_1^2)]$ would not be
of the form described.

\begin{proposition}
  \label{prop:cnotcount:1}
  An operator $u \in SU(4)$ can be simulated using one {\tt CNOT} gate
  and arbitrary one-qubit gates from $SU(2)$ iff $\chi[\ff(u)] =
  (x+i)^2(x-i)^2$.
\end{proposition}
\begin{proof}
  $u$ is simulated using one {\tt CNOT} gate iff $u \equiv
  C_1^2$ or $u \equiv C_2^1$.  Now $\ff(C_2^1) = -i \sigma^z \otimes \sigma^x$;
  also $\ff(C_1^2) = -i \sigma^x \otimes \sigma^z$.
  Each has $\chi$ given by $(x+i)^2(x-i)^2$.
\end{proof}

In particular, we see that $C_1^2 \equiv C_2^1$. This can also be
seen from the well-known identity $(H \otimes H) C_1^2 (H \otimes
H) = C_2^1$. We will use this fact for the final case, $m=2$.

\begin{proposition}
  \label{prop:cnotcount:2}
  An operator $u \in SU(4)$ can be simulated using two {\tt CNOT}
  gates and arbitrary one-qubit gates from $SU(2)$ iff $\chi[\ff(u)]$
  has all real coefficients, which occurs iff $\mbox{tr}[\ff(u)]$
  is real.
\end{proposition}
\begin{proof}
  Since $C_1^2 \equiv C_2^1$, it is clear that $u$ can be
  simulated using two {\tt CNOT} gates iff $u
  \equiv C_1^2 (a \otimes b) C_1^2$. We decompose
  $a = R_x(\alpha) R_z(\delta) R_x(\beta)$
  decomposition and $b = R_z(\theta) R_x(\phi) R_z(\psi)$,
  and pass $R_x$ gates and $R_z$ gates outward through
  the target and control of the {\tt CNOT}
  gates. Thus we are left with $u \equiv C_1^2 [R_z(\delta) \otimes
  R_x(\phi)] C_1^2$. Explicit computation yields
  $\chi[\ff(C_1^2 [R_z(\delta) \otimes R_x(\phi)] C_1^2)] =
  (x + e^{i(\delta+\phi)})(x+e^{-i(\delta+\phi)})(x + e^{i(\delta
  -\phi)})(x + e^{-i(\delta - \phi)})$. On the other hand, if
  $\chi[\ff(u)]$ has all real coefficients, then the eigenvalues
  come in conjugate pairs; it follows from this and Proposition
  \ref{prop:invariants} that $\chi[\ff(u)]$ is as above for some
  $\delta, \phi$.

  Finally, we note that for $u \in SU(N)$, and $\chi(u) =
  \prod (x - \lambda_i)$, we have $\prod \lambda_i = 1$. Thus
  $\chi(u) = \left(\prod \overline{\lambda_i}\right)
  \prod (x-\lambda_i) =
  \prod (\overline{\lambda_i} x - 1)$. It follows that the
  coefficient of $x^k$ is the complex conjugate of the coefficient
  of $x^{N-k}$. In particular, for $N=4$, the coefficient of $x^2$
  is real and the coefficients of $x^3, x$ are $\tr(u)$ and its
  conjugate. Since the constant term and the $x^4$ coefficient are
  $1$, we see $\chi(u)$ has all real coefficients iff
  $\tr(u)$ is real.
\end{proof}

\section{Synthesis Algorithm and Its Validation}
\label{sec:examples}

The results of Section \ref{sec:CNOT}
can be combined with the techniques of Propositions
\ref{prop:cnotcount:2} and \ref{prop:invariants} and the published
literature to yield an explicit circuit synthesis algorithm:

\begin{itemize}
  \item Given the matrix of a unitary operator $u \in U(4)$,
  divide it by $\sqrt[4]{\det(u)}$ to ensure $u \in SU(4)$.
  \item Compute $\chi[\ff(u)]$ to determine whether $u$ requires zero,
  one, two, or three {\tt CNOT} gates.
  \item If $u$ requires zero or one {\tt CNOT} gates, use the techniques of the
  proof of Proposition \ref{prop:invariants1} to determine which
  one-qubit operators are required.
  \item If $u$ requires two {\tt CNOT} gates,
  find the roots of $\chi[\gamma(u)]$ and
  determine the $\delta,\phi$ of Proposition \ref{prop:cnotcount:2}.
  Then use the methods of Proposition \ref{prop:invariants1}
  to determine what one-qubit gates are required at the ends of the circuit.
  \item Finally, if $u$ requires three {\tt CNOT} gates, apply the
  methods of the literature \cite{ShendeEtAl:04}.
\end{itemize}

By construction, the algorithm produces {\tt CNOT}-optimal circuits
in all cases.  It also outperforms those in
\cite{BullockM:03,ShendeEtAl:04,VidalDawson:03, VatanWilliams:03}
in important special cases, as shown below.

  \begin{example}\label{ex:hth}
     Many quantum algorithms, notably Grover's quantum search \cite{Grover97}
     and Shor's number factoring \cite{Shor97}, use the operator
     $u = H \otimes H$ to create superpositions. Computing $\gamma(u)$
     allows our synthesis algorithm to recognize that
     $u$ admits a quantum circuit containing no {\tt CNOT}s.
  \end{example}

    This example is less trivial than it seems: while writing
    $u = H \otimes H$ makes it obvious that $u$ requires no {\tt CNOT} gates,
    a synthesis procedure will not receive an input of $u = H \otimes H$ 
    but rather of the $4\times 4$ matrix corresponding to $u$. It is not
    {\em a priori} clear that any worst-case {\tt CNOT}-optimal circuit
    decomposition will implement $u$ without {\tt CNOT} gates.
    However, several previously published algorithms do.
    For the next example, previous two-qubit synthesis 
    techniques produce circuits with more {\tt CNOT}s than necessary.

  \begin{example}
     The operator $u$ that swaps $\ket{00} \leftrightarrow \ket{01}$ while
     fixing $\ket{10}$ and $\ket{11}$ plays a prominent role in the
     Deutsch-Josza algorithm \cite{DJ92,NielsenC:00}.
     Note that $C_2^1 (I\otimes \sigma_x)$ simulates $u$.
     Computing $\gamma(\mbox{e}^{i\pi /4}u)$ reveals that
     $u$ requires only one {\tt CNOT}.  However,
     depending on certain algorithmic choices, anywhere from one
     to four one-qubit gates could appear. In any event, this compares favorably
     to previous work \cite{ShendeEtAl:04} which synthesizes a circuit
     with two {\tt CNOT} and five one-qubit gates.
   \end{example}

  The algorithmic choices mentioned above come in two flavors.
  First, as the two {\tt CNOT} gates $C_1^2$ and $C_2^1$ differ
  only by one-qubit gates, they are equivalent from the
  perspective of our methods. However, the number of one-qubit
  gates present in the resulting circuit depends on which of these
  is chosen. This is a finite problem: at most three {\tt CNOT}
  gates appear and thus there are at most $8$ possibilities, so we
  simply run through them all. Additional degrees of freedom
  arise in finding a circuit that computes a given $v$ using a given $u$
  and one-qubit operators, when this is possible. The proof
  proof of Proposition \ref{prop:invariants1} describes an algorithm
  for this, and requires picking a basis of eigenvectors for a
  certain matrix. If the eigenvalues are distinct, the only degree of
  freedom is the ordering of the basis of eigenvectors
  ($4! = 24$ possibilities). However, repeated eigenvalues
  allow more flexibility in choosing basis vectors, and potentially
  non-trivial circuit optimizations.

   \begin{example}
     At the heart of Shor's factoring algorithm \cite{Shor97}
     is the Quantum Fourier Transform \cite{NielsenC:00}.
     On two qubits, it is given by the following matrix.
     \[
     \mathcal{F} = \frac{1}{2}
     {\small \left(
     \begin{array}{cccc}
       1 & 1 & 1 & 1 \\
       1 & i & -1 & -i \\
       1 & -1 & 1 & -1 \\
       1 & -i & -1 & i \\
     \end{array}
     \right)}
     \]
     Explicit computation of $\chi[\gamma(\mathcal{F})]$ reveals that
     two {\tt CNOT} gates do not suffice to simulate $\mathcal{F}$.
     Thus, the following circuit to compute $\mathcal{F}$ is {\tt CNOT}-optimal:

     
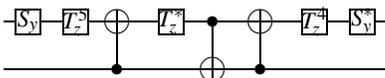
\begin{center}
       \begin{picture}(16,4)
          \put(0,0){\hWire}
          \put(0,2){\boxGate{$S_y$}}

          \put(2,0){\hWire}
          \put(2,2){\boxGate{$T_z^5$}}

          \put(4,0){\botCNOT}

          \put(6,0){\hWire}
          \put(6,2){\boxGate{$T_z^*$}}

          \put(8,0){\topCNOT}
          \put(10,0){\botCNOT}
          \put(12,0){\hWire}
          \put(12,2){\boxGate{$T_z^4$}}
          \put(14,0){\hWire}
          \put(14,2){\boxGate{$S_y^*$}}
       \end{picture}
     \end{center}

     Above, $T_z = e^{-i\sigma^z \pi/8}$ and $S_y = e^{-i\sigma^y \pi/4}$.
     Note that this circuit requires only three one-qubit gates,
     although two of these have been broken up for clarity.
     Finally, given that this circuit is {\tt
     CNOT}-optimal, it is not difficult to check by hand that its
     basic-gate count cannot be improved.
   \end{example}

\section{Timing a Hamiltonian to Compute CNOT}

Our numerical tests facilitate a new application. Given a
Hamiltonian that can be timed to compute a {\tt CNOT} modulo
one-qubit gates, we illustrate finding the correct duration. Our
example is a perturbation of $\sigma^x \otimes \sigma^x$ by
non-commutative one-qubit noise.
\begin{equation*}
H_{42}= (0.42) I \otimes \sigma^z + \sigma^x \otimes \sigma^x
\end{equation*}
Recall that a {\tt CNOT} can be constructed using one-qubit gates
and some time-iterate of the Hamiltonian $\sigma^x \otimes
\sigma^x$. However, to handle the noise term, existing techniques
resort to Trotterization, which implements $exp(A+B)$ by
separately turning on $A$ and $B$ for short periods of time. Below
we find a simpler, direct implementation of {\tt CNOT} from
$H_{42}$. It is especially interesting in light of concerns about
the scalability of Trotterization \cite{ChildsHN:03}.

 We compute $\gamma(\mbox{e}^{i H_{42}t})$
for uniformly-spaced trial values of $t$ and seek out those values
at which the characteristic polynomial nears
$p(x)=(x^2+1)^2=x^4+2x^2+1$. Our implementation in {\tt C++}
finds $t_{\tt CNOT}=0.80587$ in twenty seconds on a common
workstation. Hence, we produce a {\tt CNOT} from $H_{42}$
and one-qubit gates without Trotterization.  Specifically, since
$\mbox{e}^{iH_{42}t_{\tt CNOT}}$ implements $C_2^1$ up to one-qubit
operators, we use the technique of Proposition \ref{prop:invariants1}
to compute the relevant one-qubit operators. We find that the matrices
\begin{equation*}
\begin{array}{rr}
a_2=\frac{1}{2} \left(
\begin{array}{rr}
1-i & -1+i \\
1+i & 1+i \\
\end{array}
\right) & c_2=0.707107 \left(
\begin{array}{rr}
-1 & -1 \\
1 & -1 \\
\end{array}
\right)
\end{array}
\end{equation*}
\begin{equation*}
b_2= \left(
\begin{array}{rr}
-0.21503-0.976607i & 0 \\
0 & -0.21503+0.976607i \\
\end{array}
\right)
\end{equation*}
\begin{equation*}
d_2= \left(
\begin{array}{rr}
0.152049+0.690566i & 0.690566-0.152049i \\
-0.690566-0.152049i & 0.152049-0.690566i \\
\end{array}
\right)
\\
\end{equation*}
satisfy $C_2^1=(a_2 \otimes b_2) \mbox{e}^{iH_{42}t_{\tt CNOT}}
(c_2 \otimes d_2)$ with numerical precision of $10^{-6}$.



Further numerical experiments suggest that building a {\tt CNOT}
is possible whenever $0.42$ is replaced by a weight $w$, $0 \leq w \leq 1$.
However, we have no analytical proof of this. Numerical experiments also
suggest the {\em impossibility} of timing the Hamiltonian
$H_{XYZ}=\sigma^x \otimes \sigma^x + \sigma^y \otimes \sigma^y +
\sigma^z \otimes \sigma^z$ so as to compute a {\tt CNOT}.
In other words, trying values of $t$ in the range $-10 \leq t \leq 10$
as above produced no candidate durations.


\section{Conclusions and Future Work} \label{sec:conclusions}

   Our work addresses small-circuit structure in two-qubit unitary operators.
  In particular, we contribute tests for such structure, and our techniques
  can be viewed as algorithms for finding small circuits when they exist.
  We detail such an algorithm that produces the minimal possible number
  of {\tt CNOT} gates (zero, one, two or three) {\em for each input}.
  It is illustrated on circuit examples derived from well-known applications.

  The one-{\tt CNOT} test has an additional use.
  It provides a numerical method for timing a given
  two-qubit Hamiltonian $H$ so that $\mbox{e}^{i t H}$ realizes
  a {\tt CNOT} gate up to local unitary operators
  (one-qubit gates,) given this is possible for $H$.

     {\bf Acknowledgments.} This work is supported by the
   DARPA QuIST program and an NSF grant. The views and
   conclusions contained herein are those of the authors
   and should not be interpreted as neces\-sarily representing official
   policies or endorsements of employers and funding agencies.

\section*{Appendix}
\label{sec:delta2}

\begin{proposition} \label{prop:invariants1}
Let $\ff: SU(4) \to SU(4)$ be given by the formula $u \mapsto u
(\sigma^y)^{\otimes 2} u^t (\sigma^y)^{\otimes 2}$. Then for $u, v
\in SU(4)$, $u \equiv v \iff \chi[\gamma(u)] = \chi[\pm
\gamma(v)]$.
\end{proposition}

\begin{proof}
  By definition, $u \equiv v \iff u = (a \otimes b)\lambda v (a'
  \otimes b')$ for some one-qubit operators $a,b,a',b'$ and some
  scalar $\lambda$. Requiring $u,v \in SU(4)$ implies $\lambda =
  \pm 1, \pm i$. We show below that $u = (a \otimes b)v(a' \otimes b')
  \iff \chi[\gamma(u)] = \chi[\gamma(v)]$; the proposition
  then follows from the fact that $\gamma(iu) = -\gamma(u)$.

  We recall that there exist $E \in SU(4)$ such that $E~SO(4)~E^*
  = SU(2)^{\otimes 2} =
  \{a \otimes b: a,b \in SU(2)\}$. Such matrices are characterized by the property
  that $EE^t = -\sigma^y \otimes \sigma^y$. This and related issues
  have been exhaustively dealt with in several papers
  \cite{BennettEtAl:96,HillWooters:97,
  KhanejaBG:01a,
  LewensteinEtAl:01}.

  The property $\chi[\ff(u)] = \chi[\ff(v)]$ is not changed by replacing
  $\ff$ with $E^* \ff E$. Using the fact $\sigma^y \otimes \sigma^y = EE^t =
  (EE^t)^*$, we compute:
  $E^* \ff(u) E = E^* u E E^t u^t E^{t*} E^* E
  = (E^* u E)(E^* u E)^t$

  By making the substitution $u \mapsto EuE^*$; it
  suffices to prove: for $u, v \in SU(4)$, there
  exists $x, y \in SO(4)$ such that $xuy = v$ iff $\chi[uu^t] =
  \chi[vv^t]$.  Here, $SO(4)$ is the real matrices within $SU(4)$.

  Note that for $P$ symmetric unitary, $P^{-1} =
  \overline{P}$, hence $[P+\overline{P}, P-\overline{P}]=0$. It
  follows that the real and imaginary parts of $P$ share an
  orthonormal basis of eigenvectors. As they are moreover real
  symmetric matrices, we know from the spectral theorem that their
  eigenvectors can be taken to be real.
  Thus there exists $q \in SO(4)$ such
  that $quu^t q^*$ is diagonal. By re-ordering (and negating)
  the columns of $q$, we can re-order the diagonal elements of
  $quu^t q^*$ as desired. Thus if $\chi[uu^t]=\chi[vv^t]$,
  we can find $q, r \in SO(4)$ such that
  $quu^t q^t = r vv^t r^t$ by diagonalizing both; then
  $(v^* r^t q u)(v^* r^t q u)^t = I$. Let $s = v^* r^t q u \in
  SO(4)$. We have $ q^t r v s = u$, as desired.
\end{proof}

\end{document}